\pdfoutput=1
\PassOptionsToPackage{numbers,sort&compress}{natbib}
\documentclass[3p,times,a4paper]{elsarticle}

\usepackage[T1]{fontenc}

\usepackage{amsmath}
\usepackage{graphicx}
\usepackage{booktabs}
\usepackage{tabularx}
\usepackage{array}
\usepackage{algorithm}
\usepackage{algpseudocode}
\usepackage{listings}
\usepackage{xcolor}
\usepackage{microtype}
\usepackage[hidelinks]{hyperref}
\hypersetup{pdftitle={FedReason: Large Language Models as Weight-Space Meta-Aggregators for Heterogeneous Federated Learning},
  pdfauthor={Ali Mohammadi Ruzbahani, Abbas Yazdinejad, Hadis Karimipour}}

\graphicspath{{figures/}}

\newcommand{\code}[1]{\texttt{\detokenize{#1}}}
\newcommand{\pmstd}[1]{\,{\scriptsize$\pm$#1}}
\newcommand{\binv}{b_{\mathrm{inv}}}
\newcommand{\bprog}{b_{\mathrm{prog}}}
\newcolumntype{Y}{>{\raggedright\arraybackslash}X}

\begin{document}

\begin{frontmatter}

\title{FedReason: Large Language Models as Weight-Space Meta-Aggregators for Heterogeneous Federated Learning}

\author[ucal]{Ali Mohammadi Ruzbahani\corref{cor1}}
\ead{ali.mohammadiruzbaha@ucalgary.ca}
\author[ureg]{Abbas Yazdinejad}
\ead{Abbas.Yazdinejad@uregina.ca}
\author[ucal]{Hadis Karimipour}
\ead{hadis.karimipour@ucalgary.ca}

\cortext[cor1]{Corresponding author.}

\affiliation[ucal]{organization={Smart Cyber-Physical Systems (SCPS) Lab, University of Calgary},
  city={Calgary}, state={Alberta}, country={Canada}}
\affiliation[ureg]{organization={DCAILab, University of Regina},
  city={Regina}, state={Saskatchewan}, country={Canada}}

\begin{abstract}
Federated learning servers combine client models with a fixed aggregation rule, and each rule in common use is designed for one kind of heterogeneity. We study whether a large language model (LLM) that inspects the client models can choose the rule itself, without any training on the server. In FedReason, the LLM is called every ten rounds. It analyses the weights by running code in a restricted server-side sandbox, sees only the printed results, may score a few candidates on a 500-image proxy set, and submits a short Python aggregation program. The server applies the program until the next call and falls back to FedAvg whenever its output is invalid. Full weight tensors never leave the server. We evaluated FedReason with Claude Opus 5.5 on CIFAR-10 and CIFAR-100 with ResNet-18 under Dirichlet label skew, width-heterogeneous sub-models and Byzantine clients, in 168 runs against 14 baselines. Across 12 paired label-skew runs it beat FedAvg by 1.26 percentage points on average (10 wins), and a control that spends the same per-round proxy budget on choosing among standard rules by 1.08 points. It was the strongest server-side rule on CIFAR-100 but did not beat SCAFFOLD, which changes client training. With three of ten clients malicious, it excluded all sign-flipping clients from the first round and stayed about one point below FLTrust without training a root model. Without proxy feedback, accuracy fell from 73.8\% to 38.8\% in one seed. A 60-round run cost 0.79 USD in API fees on average.
\end{abstract}


\end{frontmatter}

\section{Introduction}
\label{sec:intro}

This paper asks whether a large language model (LLM) that reads the client models on a federated learning (FL) server can decide how to combine them, and do so better than a fixed aggregation rule. We test this when clients differ in their label distributions, in the width of model they can train, or in whether they are honest. Each standard aggregation rule encodes one assumption about what goes wrong, and a deployment rarely knows in advance which assumption holds. We add one constraint that keeps the comparison fair: the server trains nothing. It may evaluate a model on a small labelled proxy set a fixed number of times, but it may not fit any parameter to that set by gradient descent.

\begin{figure}[t]
  \centering
  \includegraphics[width=\textwidth]{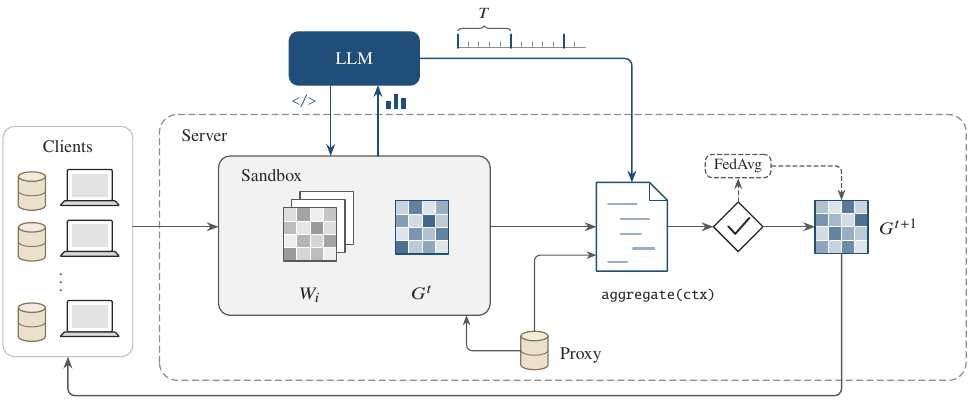}
  \caption{Overview of FedReason (dashed line: server boundary).}
  \label{fig:overview}
\end{figure}

Our idea is to let the LLM write the aggregation rule as a program. In every round of FL the server holds a set of freshly trained models of one architecture, and these models carry information that a weighted average throws away. The rows of a client's classifier show which classes the client has seen. The size and direction of its update show whether it agrees with the other clients, and a poisoned update usually stands out on both counts. In FedReason (Fig.~\ref{fig:overview}), the server calls the LLM every $T$ rounds. The LLM examines the weights by running Python code in a restricted sandbox on the server and reads only what that code prints. It may score a few candidate models on the proxy set. It then submits a short function, \code{aggregate(ctx)}, which the server executes in every round until the next call.

A program can express whichever rule the evidence supports: sample-weighted averaging, clipping of large updates, exclusion of clients that look malicious, separate treatment of the classifier rows, or a server step with momentum. When the evidence changes, the LLM replaces it. Since the LLM looks at the evidence before it chooses, the server does not have to know in advance whether its clients are skewed, limited in compute or malicious. The program keeps running between calls, so a 60-round run needs six calls, the API cost is bounded and most rounds incur no LLM latency. The server checks the program's output in every round and falls back to federated averaging (FedAvg) when the output is invalid, so a faulty program cannot stop training. Each decision is stored as code together with the rationale the LLM gave for it and can be audited after training. The raw weight tensors stay on the server, and the clients run ordinary local training.

FL trains a shared model on data that remain on the clients~\cite{mcmahan2017communication,kairouz2021advances}. FedAvg weights each client model by its sample count~\cite{mcmahan2017communication}, and most alternatives replace this formula by another one that is fixed before training starts and designed for one problem, such as label skew~\cite{karimireddy2020scaffold,reddi2021adaptive}, sub-models of different widths~\cite{diao2021heterofl} or malicious clients~\cite{blanchard2017krum,cao2021fltrust}. Methods that adapt during training tune the client weights or pick one rule from a fixed menu~\cite{wu2021fedadp,haque2025fedstrategist}. Recent LLM agents for FL design whole systems or choose among rules from specifications, data characteristics, client profiles and scalar metrics~\cite{li2025helmsman,wu2026flmaestro,pang2026automating}. None of them looks at the weights. Takehi et al.~\cite{takehi2026sparks} recently reported, in a blog post that has not been peer reviewed, that a frontier LLM could tell from the weights of a small policy network which game it had been trained on. The LLM could also write or edit small networks without gradient training. The authors name model merging beyond weight averaging as a possible use. FedReason puts this suggestion to a controlled test in FL.

We make four contributions.
\begin{itemize}
  \item We formulate meta-aggregation, in which the server chooses the aggregation operator during training as a sequence of programs written from the observed client models, under budgets on LLM cost and proxy evaluations and without server-side training. It generalises per-round selection among fixed rules~\cite{haque2025fedstrategist} (Section~\ref{sec:problem}).
  \item We present FedReason, in which a tool-using LLM inspects the client models in a server-side sandbox and writes an aggregation program that the server checks in every round, falling back to FedAvg when the check fails. The clients are unchanged and every API call is logged (Section~\ref{sec:method}).
  \item We evaluate FedReason in 168 runs on CIFAR-10 and CIFAR-100 under label skew, width heterogeneity and two Byzantine attacks, against 14 baselines that include a control spending the same per-round proxy budget on selecting among standard rules. FedReason beats FedAvg in 10 of 12 paired label-skew runs and is the most accurate server-side rule on CIFAR-100, but it does not beat SCAFFOLD. Under attack it comes within about one point of FLTrust without training on the server (Sections~\ref{sec:setup} and~\ref{sec:results}).
  \item We analyse what the LLM does, through ablations that withhold the weights or the proxy feedback, an audit of all 120 programs accepted in the main runs, and an account of the cases in which its inferences were wrong (Sections~\ref{sec:results} and~\ref{sec:discussion}).
\end{itemize}
Section~\ref{sec:related} reviews related work, Section~\ref{sec:problem} states the problem, Section~\ref{sec:method} describes FedReason, Sections~\ref{sec:setup} and~\ref{sec:results} report the experiments, Section~\ref{sec:discussion} discusses limitations and Section~\ref{sec:conclusion} concludes.

\section{Related work}
\label{sec:related}

When client data are not identically distributed, the local optima differ and the average of the client models moves away from the optimum of the global objective. The effect grows with label skew and with the amount of local computation~\cite{hsu2019measuring,li2022niid}, and Dirichlet label partitions with a concentration parameter are the common way to control label skew in experiments~\cite{hsu2019measuring,li2022niid}. Some remedies change local training. FedProx adds a proximal term~\cite{li2020fedprox}, and SCAFFOLD corrects client drift with control variates, at the cost of a second vector sent by every client~\cite{karimireddy2020scaffold}. Others change only the server: FedAvgM adds server momentum~\cite{hsu2019measuring}, FedAdam applies an adaptive optimiser to the averaged update~\cite{reddi2021adaptive} and FedNova normalises updates by the number of local steps~\cite{wang2020tackling}. A third line of work adapts the client weights during training. FedAdp derives them from the angle between each local update and the global update~\cite{wu2021fedadp}. FedLAW learns them together with a global shrinking factor by gradient descent on proxy data held by the server~\cite{li2023fedlaw}, and FedAWA optimises them from the client update vectors without extra data~\cite{shi2025fedawa}. FedStrategist selects, in every round, one rule from a fixed set of robust aggregators with a contextual bandit driven by diagnostics of the updates~\cite{haque2025fedstrategist}. These methods either fix the form of the rule before training, so that what adapts is at most a set of client weights or the state of a server optimiser, or pick from a fixed menu. FedReason searches a larger space, because a program can treat classifier rows separately, clip or exclude clients and change the server step over time. It chooses within that space by analysis and a few forward evaluations rather than by gradient descent. FedLAW is its closest gradient-based counterpart, since it also relies on server-side proxy data.

When clients cannot train the full model, they can train sub-models extracted from it. HeteroFL gives each client a fixed set of leading channels in every layer, averages each coordinate over the clients that cover it, and evaluates with static batch normalisation (sBN), which recomputes the normalisation statistics after training~\cite{diao2021heterofl}. FedRolex rolls the extraction window across rounds so that all channels are trained evenly~\cite{alam2022fedrolex}. Both average coordinate-wise with fixed weights. FedReason keeps HeteroFL's static extraction, so any difference between the two comes from aggregation.

Malicious clients can poison FL through crafted model updates~\cite{fang2020local} or through flipped training labels~\cite{tolpegin2020data}. Krum selects the update closest to its neighbours~\cite{blanchard2017krum}. The coordinate-wise median and the trimmed mean bound the influence of a minority of corrupted clients, the trimmed mean given an upper bound on their fraction~\cite{yin2018byzantine}. FLTrust keeps a small root dataset on the server and trains a server update on it in every round. It scales each client update to the norm of that server update and weights it by the rectified cosine similarity between the two~\cite{cao2021fltrust}. Robust aggregation rules assume that honest updates resemble each other, which label skew breaks~\cite{li2019rsa}, so they can discard useful honest updates. FedReason uses the same kind of server data as FLTrust, but only for forward evaluations. A different answer to untrusted parties is to stop sharing parameters altogether. In our recent work, heterogeneous peers improve their models by exchanging privacy-preserving queries, and a Byzantine-resilient committee filters the answers before they are used~\cite{ruzbahani2026trustless}. FedReason targets the more common setting in which the server does receive the client models and has to decide how to combine them.

LLMs can improve candidate solutions from scored feedback. OPRO proposes new solutions from a history of scored ones~\cite{yang2024opro}, and FunSearch pairs LLM-written programs with an automatic evaluator to find new constructions~\cite{romeraparedes2024funsearch}. FedReason follows the same pattern of proposal and evaluation, with a budgeted proxy set as evaluator and programs that act on model weights. LLM agents have also been built for FL. Helmsman synthesises a complete FL system from a specification with collaborating agents and refines it in a sandboxed simulation before deployment~\cite{li2025helmsman}. FL-MAESTRO uses agents at run time to set the communication topology, per-client resources and the aggregation rule~\cite{wu2026flmaestro}, and Pang et al. let an LLM infer a suitable aggregation strategy from user-provided or detected data characteristics~\cite{pang2026automating}. These systems act on specifications, data descriptions, client profiles or scalar metrics. Methods that edit trained weights directly do exist outside FL. Approximate machine unlearning, for example, removes the influence of selected training data with a single Newton-style update, and our TensorUnlearn makes the required curvature tractable for wide layers by compressing Kronecker-factored estimates with tensor networks~\cite{ruzbahani2026tensorunlearn}. In such methods the edit follows a fixed formula. The report of Takehi et al.~\cite{takehi2026sparks} suggests that an LLM can decide the edit itself by working on the weights. To our knowledge, no earlier FL method lets an LLM read the client models during training and write the aggregation code. Our metrics-only ablation measures what that access adds.

\section{Problem formulation}
\label{sec:problem}

We consider $K$ clients $i\in\{0,\dots,K-1\}$. Client $i$ holds a private dataset $\mathcal{D}_i$ of $n_i$ samples with labels $y\in\{0,\dots,C-1\}$, and we write $N=\sum_in_i$ and $p_i=n_i/N$. A model $f(\cdot\,;w)$ with parameters $w\in\mathbb{R}^d$ is trained to minimise
\begin{equation}
  F(w)=\sum_{i}p_iF_i(w),\qquad
  F_i(w)=\frac{1}{n_i}\sum_{(x,y)\in\mathcal{D}_i}\ell\big(f(x;w),y\big),
\end{equation}
where $\ell$ is the cross-entropy loss. Training runs for $R$ rounds. In round $t$ the server sends the global model $G^t$, or a sub-model $\mathcal{E}(G^t;\mathcal{I}_i)$ selected by an index map $\mathcal{I}_i$, to client $i$, which trains it locally and returns $W_i^t$. We write $U_i^t=W_i^t-\mathcal{E}(G^t;\mathcal{I}_i)$ for the client's update and $m_i^t$ for its metadata (sample count, number of local steps, mean training loss and width). The server then applies an aggregation operator
\begin{equation}
  G^{t+1}=\mathcal{A}_t\big(G^t,\{W_i^t\},\{m_i^t\}\big).
  \label{eq:aggregation}
\end{equation}
FedAvg uses $\mathcal{A}^{\mathrm{avg}}=\sum_ip_iW_i^t$ in every round. Server optimisers such as FedAvgM and FedAdam, adaptive weightings such as FedAdp, and robust rules such as the median, the trimmed mean and Krum are instances of Eq.~\eqref{eq:aggregation}, some of them with internal state.

Under model heterogeneity, client $i$ trains a network whose layers keep a fraction $r_i\in\{1,\frac12,\frac14\}$ of the channels of the global model. The index map takes the leading channels of each layer (static extraction, as in HeteroFL) or a window that shifts from round to round (rolling extraction, as in FedRolex). Scattering $W_i^t$ back into the coordinates of the full model gives $\hat W_i^t\in\mathbb{R}^d$ and a binary coverage mask $M_i^t$. A coordinate-wise aggregation with client coefficients $c_i\ge0$ sets, for every coordinate $j$ covered by at least one client with $c_i>0$,
\begin{equation}
  \big[G^{t+1}\big]_j=\frac{\sum_ic_i\,[M_i^t]_j\,[\hat W_i^t]_j}{\sum_ic_i\,[M_i^t]_j},
  \label{eq:hetero}
\end{equation}
and keeps $[G^t]_j$ for the other coordinates. HeteroFL and FedRolex use $c_i=1/K$, and coordinate-wise FedAvg uses $c_i\propto n_i$. Under attack, a set $\mathcal{B}$ of clients that the server does not know is malicious. Sign-flipping clients~\cite{li2019rsa} return $G^t-s\,U_i^t$ with $s=3$, and label-flipping clients train on the labels $y\mapsto C-1-y$~\cite{fang2020local}.

The server holds a small, clean labelled proxy set $\mathcal{P}$, as FLTrust assumes~\cite{cao2021fltrust}; in our experiments $\mathcal{P}$ is class-balanced and disjoint from all client data. A forward evaluation $\mathrm{ev}_{\mathcal{P}}(G)$ returns the accuracy, the loss and the per-class accuracy of a model $G$ on $\mathcal{P}$. The test set is never visible to any aggregator.

A meta-aggregator $\mathcal{M}$ chooses the operator during training. At invocation rounds $t_1=1<t_2<\dots<t_J$ it observes
\begin{equation}
  \mathcal{O}_{t_j}=\big(\mathcal{C}^{t_j},\,\mathcal{H}^{t_j},\,\nu_{j-1},\,\pi_{j-1}\big),\qquad
  \mathcal{C}^{t}=\big(t,\,G^t,\,\{W_i^t\},\,\{U_i^t\},\,\{m_i^t\}\big),
\end{equation}
that is, the round context $\mathcal{C}^t$, the history $\mathcal{H}$ of proxy metrics and aggregation coefficients, its own notes $\nu_{j-1}$ from the previous invocation and the previous program $\pi_{j-1}$. It outputs a program $\pi_j\in\Pi$, where $\Pi$ is the set of programs that map a round context to a model with the keys and tensor shapes of $G^t$, and the server applies $G^{t+1}=\pi_j(\mathcal{C}^t)$ for $t_j\le t<t_{j+1}$, with $t_{J+1}=R+1$. The meta-aggregation problem is
\begin{equation}
  \max_{\mathcal{M}}\ \mathbb{E}\big[\mathrm{Acc}_{\mathrm{test}}(G^{R+1})\big]
  \quad\text{s.t.}\quad
  \sum_{j=1}^{J}\kappa_j\le B,\qquad
  e_j^{\mathrm{inv}}\le\binv\ \ \forall j,\qquad
  e_t^{\mathrm{prog}}\le\bprog\ \ \forall t,
  \label{eq:meta}
\end{equation}
where $\kappa_j$ is the API cost of invocation $j$, $e_j^{\mathrm{inv}}$ the number of proxy evaluations used during invocation $j$ and $e_t^{\mathrm{prog}}$ the number used by the active program in round $t$. The expectation is over partitions, local training and the randomness of $\mathcal{M}$. Two further constraints are structural: neither $\mathcal{M}$ nor any program may run gradient-based optimisation on the server, and full weight tensors are never sent to an external model provider.

The test accuracy is never observed, so $\mathcal{M}$ has to decide from weight-space evidence and a few proxy evaluations. The set $\Pi$ contains every purely server-side baseline of this study (FedAvg, FedAvgM, FedAdam, FedNova, FedAdp, the median, the trimmed mean, Krum and the HeteroFL average), because each is a short program over $\mathcal{C}^t$ and a persistent state. An oracle meta-aggregator would therefore match the best of these rules in every setting. FedProx and SCAFFOLD lie outside $\Pi$ because they change what the clients compute, and FLTrust lies outside it because it trains on the server.

\section{The FedReason framework}
\label{sec:method}

FedReason is a meta-aggregator for problem~\eqref{eq:meta} that changes only the server (Fig.~\ref{fig:overview} and Algorithm~\ref{alg:fedreason}). FedAvg is the active rule until the first program is accepted. In every round with $(t-1)\bmod T=0$, the server opens a new tool-use conversation with the LLM, loads the round context into an analysis sandbox and lets the LLM investigate it until it submits a program that passes a dry run on the current round. That program aggregates the current round and every later round until the next invocation. Because the first invocation happens in round~1, before the first aggregation, every round is aggregated by an LLM-written program unless a program fails. Between invocations there are no API calls.

\begin{algorithm}[!t]
\caption{FedReason, server side.}
\label{alg:fedreason}
\begin{algorithmic}
\Require initial model $G^1$, rounds $R$, period $T$, budgets $\binv$ and $\bprog$, proxy set $\mathcal{P}$
\State $\pi\gets$ FedAvg;\ \ $\mathcal{H}\gets\emptyset$;\ \ $\nu\gets\emptyset$
\For{$t=1,\dots,R$}
  \State every client $i$ trains locally and returns $W_i^t$;\ \ $U_i^t\gets W_i^t-\mathcal{E}(G^t;\mathcal{I}_i)$
  \State $\mathcal{C}^t\gets\big(t,G^t,\{W_i^t\},\{U_i^t\},\{m_i^t\}\big)$
  \If{$(t-1)\bmod T=0$}
    \State $(\pi',\nu')\gets\textsc{Invoke}(\mathcal{C}^t,\mathcal{H},\nu,\pi)$
    \If{$\pi'\neq\bot$}
      \State $\pi\gets\pi'$;\ \ $\nu\gets\nu'$
    \EndIf
  \EndIf
  \State $\tilde G\gets\pi(\mathcal{C}^t)$ \Comment{at most $\bprog$ proxy evaluations}
  \If{$\textsc{Valid}(\tilde G,G^t)$}
    \State $G^{t+1}\gets\tilde G$
  \Else
    \State $G^{t+1}\gets$ FedAvg$(\mathcal{C}^t)$; keep the error for the next brief
  \EndIf
  \State $\mathcal{H}\gets\mathcal{H}\cup\big\{\big(t,\mathrm{ev}_{\mathcal{P}}(G^{t+1}),\text{coefficients}\big)\big\}$
\EndFor
\State \Return $G^{R+1}$
\Statex
\Function{Invoke}{$\mathcal{C}^t,\mathcal{H},\nu,\pi$}
  \State load $\mathcal{C}^t$ and $\mathcal{H}$ into the sandbox; send the brief built from $\mathcal{C}^t$, $\mathcal{H}$, $\nu$ and $\pi$
  \For{at most ten model requests}
    \If{the LLM calls \code{python}}
      \State run the code and return its printed output \Comment{at most $\binv$ proxy evaluations}
    \ElsIf{the LLM calls \code{submit_policy} with program $\pi'$ and notes $\nu'$}
      \If{$\textsc{Valid}\big(\pi'(\mathcal{C}^t),G^t\big)$} \Comment{dry run}
        \State \Return $(\pi',\nu')$
      \EndIf
      \State return the error to the LLM \Comment{at most three submissions}
    \EndIf
  \EndFor
  \State \Return $(\bot,\nu)$
\EndFunction
\end{algorithmic}
\end{algorithm}

\subsection{Reading the weights in a server-side sandbox}
\label{sec:sandbox}

During an invocation the LLM has two tools: \code{python}, which runs code in a persistent namespace on the server and returns what it prints, and \code{submit_policy}, which submits a program. The namespace holds read-only copies of the global model \code{G}, the client models \code{W}, the updates \code{U}, the metadata \code{meta} and the \code{history} of proxy metrics and coefficients. It also holds a table of layers and a description of the setting (dataset, number of classes, architecture, client widths, current round and budgets). Helper functions compute cosine similarities, sample-weighted and general weighted averages (coordinate-wise for sub-models, Eq.~\eqref{eq:hetero}), averages of single rows of a tensor with their own client coefficients (for instance per class in the classifier), the client-by-parameter update matrix and a server step $G^t+\eta(\tilde G-G^t)$. A budgeted function \code{proxy_eval} evaluates any candidate model on the proxy set, at most $\binv=6$ times per invocation.

A helper \code{summary()} prints a compact table that gives the LLM a starting point. For each client the table lists the sample count, the number of local steps and the training loss. It adds the update norm with a robust $z$-score (median and median absolute deviation) and the relative update size of each layer group (stem, four stages and classifier). A leave-one-out cosine compares the client's update with the sample-weighted mean update of the other clients, for the whole model and per layer group, and a shift score summarises the change in its normalisation statistics. The helper also prints the pairwise cosines of the updates and, for each client, the classes whose classifier bias and rows changed most. Under label skew, the classifier rows of classes that a client has not seen receive systematically different updates from the rows of classes it has seen. The label composition of a client can therefore be read from its weights, although the server never sees its data. The client label histograms, the identities of malicious clients and the test set are never placed in the namespace.

The printed output of each call is cut to 6{,}000 characters, and this output is the only view of the weights that the LLM gets. Code runs with gradients disabled, restricted built-in functions and an import whitelist (\code{numpy}, \code{torch}, \code{scipy}, \code{math} and a few standard modules). Before any code runs, a check of its syntax tree rejects private and reflective attribute access and a list of file, process and automatic-differentiation functions. While it runs, an interpreter audit hook blocks file writes, sockets and subprocesses. Errors come back as tracebacks, which lets the LLM correct its code.

\subsection{Aggregation programs and invocations}
\label{sec:programs}

A program is a Python function \code{aggregate(ctx)}, submitted together with a short rationale and with notes for the next invocation. In every round the server calls it with a context \code{ctx} that exposes the objects and helpers of the sandbox. The context also holds a dictionary \code{ctx.memory} that persists across rounds and programs (for server momentum, smoothed statistics or per-client suspicion counters) and a proxy evaluator limited to $\bprog=3$ calls per round. The program returns a model with exactly the keys and tensor shapes of $G^t$ and may report its client coefficients for logging. On submission the server dry-runs the program on the current round. If this fails, the error goes back to the LLM, which may resubmit up to three times in total. In every round the output is checked for its key set and shapes, finite values, non-negative running variances and an update norm $\lVert G^{t+1}-G^t\rVert$ of at most ten times the largest client update norm. If the program raises an exception or its output fails a check, the round uses FedAvg, \code{ctx.memory} is rolled back and the error is reported in the next brief.

Each invocation is a new conversation of at most ten model requests. Between invocations, the LLM remembers only what the server writes into the first message, which we call the brief. It contains the setting description, the client metadata, the history of proxy metrics and coefficients, and the previous program with its rationale, its per-round proxy accuracy, errors and printed diagnostics. It also carries the LLM's own notes from the previous invocation (at most 4{,}000 characters) and the remaining budgets. If the turn limit passes without an accepted program, the previous program stays active.

The system prompt is static. It defines the role of a server-side meta-aggregator, documents the interfaces and sets seven rules. The LLM must not train on the server: automatic differentiation, optimisers and any iterative fitting to proxy data are excluded, and the budgeted \code{proxy_eval} is the only forward pass. Files and the network are off limits. The prompt names expected test accuracy as the target, warns that clients may be heterogeneous, skewed or malicious, asks for cheap and numerically safe programs, and requires every invocation to end with \code{submit_policy}. It also states that accuracy on 500 proxy images is noisy by about $\pm2$ percentage points and that small differences should not decide between candidates. A short guide lists heuristics such as layer-wise update geometry, classifier-bias evidence of label skew, normalisation statistics, clipping, the median or trimmed mean, cosine-based down-weighting, server steps and momentum. A minimal example program shows the interface with a sample-weighted body, class-evidence weighting of the classifier rows and server momentum. The prompt thus suggests building blocks without prescribing any, and Section~\ref{sec:res-llm} shows that one of its suggestions can do harm.

Proxy access is a confound when FedReason is compared with methods that use no server data, since a server that can evaluate candidates could simply pick among standard rules. We therefore budget it and add a control, ProxySel, which spends the same three evaluations per round on choosing among three standard candidates (Section~\ref{sec:setup}). FedReason may also use six evaluations per invocation for analysis. When every program is accepted at its first submission, as all were in our runs, a 60-round run with six invocations may therefore request at most $6\cdot6+60\cdot3=216$ evaluations, against 180 for ProxySel; rejected dry runs could add at most 54 more. The programs used far fewer (Section~\ref{sec:res-llm}).

\subsection{Safeguards, privacy and cost}
\label{sec:privacy}

FedReason keeps the usual FL trust model, in which an honest-but-curious server receives the individual client models. Like Krum and FLTrust it needs these individual models, so it cannot be combined with secure aggregation~\cite{bonawitz2017practical}, and the server faces the same risk of reconstructing training data from updates as it does under FedAvg~\cite{zhu2019deep}. The new party is the LLM provider, which we also treat as honest-but-curious. It receives the static system prompt, the brief (server-side aggregates such as sample counts, losses, proxy metrics and coefficients, together with text that the LLM or its programs wrote earlier), the truncated outputs of the LLM's code and the server's replies to submissions. Full weight tensors never leave the server. The printed outputs can still contain individual weight values, and many statistics that the LLM asks for reveal the label composition of clients. The LLM also writes such inferences into its notes, which are sent again with later briefs. The FL server already holds this information, but FedReason shares it with one more party.

Malicious clients control their weights and the numbers they report, but they cannot inject text, because every client-dependent field of the brief is a number formatted by the server. Code written by an LLM can contain bugs and security weaknesses~\cite{pearce2022asleep}, so we treat it as untrusted: it runs in the restricted namespace, and its output is checked in every round with a FedAvg fallback. A restricted Python namespace is not a security boundary, however, and a deployment should add isolation at the operating-system level.

Every API call is logged with its token counts, the model that served it and its latency. A call with $n_{\mathrm{in}}$ uncached input tokens, $n_{\mathrm{cw}}$ cache-write tokens, $n_{\mathrm{cr}}$ cache-read tokens and $n_{\mathrm{out}}$ output tokens (thinking included) costs
\begin{equation}
  c=10^{-6}\big(n_{\mathrm{in}}\,p_{\mathrm{in}}+n_{\mathrm{cw}}\,p_{\mathrm{cw}}+n_{\mathrm{cr}}\,p_{\mathrm{cr}}+n_{\mathrm{out}}\,p_{\mathrm{out}}\big)\ \text{USD},
\end{equation}
with list prices per million tokens of $p_{\mathrm{in}}=4$, $p_{\mathrm{cw}}=5$, $p_{\mathrm{cr}}=0.20$ and $p_{\mathrm{out}}=20$ USD for the model we use~\cite{anthropic2026opus55docs}. The system prompt and the tool definitions are identical across calls and are served from the prompt cache. A run has six invocations (twelve in the ablation that calls the LLM every five rounds), each of at most ten requests plus one final request that asks for a submission. Its cost therefore grows with neither the number of model parameters nor the number of rounds between invocations. A global budget stops all API calls once the total spend of the study reaches it.

\section{Experimental setup}
\label{sec:setup}

We use CIFAR-10 and CIFAR-100~\cite{krizhevsky2009learning}. Before partitioning, a class-balanced proxy set of 500 training images (50 per class on CIFAR-10 and 5 per class on CIFAR-100) is removed from the training set and given to the server. The remaining training images are split among $K=10$ clients with the per-class Dirichlet partition of NIID-Bench~\cite{li2022niid}, with concentration $\alpha\in\{0.1,0.5\}$ as in~\cite{hsu2019measuring} and at least ten images per client. Clients use random crops with 4-pixel padding, horizontal flips and per-channel normalisation. All settings use ResNet-18~\cite{he2016deep} adapted to CIFAR, with a $3\times3$ stem, no max-pooling and batch normalisation (11.17\,M parameters on CIFAR-10 and 11.22\,M on CIFAR-100). Each client trains for two local epochs with mini-batches of 64, using SGD with learning rate 0.01, momentum 0.9 and weight decay $5\times10^{-4}$, gradient-norm clipping at 10 and bfloat16 autocasting. Training lasts 60 rounds with full participation. Table~\ref{tab:settings} lists the settings. With heterogeneous widths, three clients train the full model, three the $\times\frac12$ sub-model and four the $\times\frac14$ sub-model. In the Byzantine settings three of the ten clients, drawn from a permutation seeded by the run seed, attack from the first round. Everything is implemented in PyTorch~\cite{paszke2019pytorch}, and the runs shared one NVIDIA RTX~5090 GPU, several at a time.

\begin{table}[t]
\centering
\caption{Settings and compared methods.}
\label{tab:settings}
\small
\begin{tabularx}{\textwidth}{@{}l l l l Y c@{}}
\toprule
Group & Data & Client widths & Attack & Baselines & Seeds \\
\midrule
Label skew & CIFAR-10/100, $\alpha\in\{0.1,0.5\}$ & all $\times1$ & none & FedAvg, FedProx, FedAvgM, FedAdam, FedNova, SCAFFOLD, FedAdp, ProxySel & 3 \\
\addlinespace
Width & CIFAR-10/100, $\alpha=0.5$ & $3\times1$, $3\times\frac12$, $4\times\frac14$ & none & FedAvg, HeteroFL, FedRolex, ProxySel & 2 \\
\addlinespace
Byzantine & CIFAR-10, $\alpha=0.5$ & all $\times1$ & 3 of 10 & FedAvg, Median, Trimmed mean, Krum, FLTrust, ProxySel & 2 \\
\addlinespace
Ablation & CIFAR-10, $\alpha=0.1$ & all $\times1$ & none & four FedReason variants & 1 \\
\bottomrule
\end{tabularx}
\end{table}

The label-skew baselines are FedAvg~\cite{mcmahan2017communication}, FedProx with $\mu=0.01$~\cite{li2020fedprox}, FedAvgM with server momentum 0.9 and step 1~\cite{hsu2019measuring} and FedAdam with the CIFAR-10 values of its original paper ($\eta=0.01$, $\beta_1=0.9$, $\beta_2=0.99$, $\tau=10^{-3}$)~\cite{reddi2021adaptive}. They further include FedNova with momentum-corrected step normalisation~\cite{wang2020tackling}, SCAFFOLD with option-II control variates~\cite{karimireddy2020scaffold} and FedAdp with Gompertz parameter 5~\cite{wu2021fedadp}. Because the clients use momentum, SCAFFOLD applies its correction to the weights after each local step, as in the NIID-Bench code~\cite{li2022niid}. We checked our FedNova against its official implementation. The width-heterogeneous baselines are HeteroFL~\cite{diao2021heterofl} and FedRolex~\cite{alam2022fedrolex} with the uniform coordinate-wise average of Eq.~\eqref{eq:hetero}, and FedAvg in its coordinate-wise form. Every model in these settings, for every method, is evaluated with sBN~\cite{diao2021heterofl}: before each test or proxy evaluation the running statistics are recomputed on a fixed subset of 10{,}000 client training images. The Byzantine baselines are the coordinate-wise median and the trimmed mean with a trimming fraction of 0.3 per side~\cite{yin2018byzantine}, Krum with the true number of attackers~\cite{blanchard2017krum} and FLTrust~\cite{cao2021fltrust}. FLTrust uses our proxy set as its root dataset, and its root model takes as many local steps as the median client. These four also run in the attack-free CIFAR-10 setting with $\alpha=0.5$, which measures their cost without attack. ProxySel, our control for proxy access, forms three candidates in every round: FedAvg with server step 1.0, FedAvg with server step 1.5 and FedAvgM with momentum 0.9. It keeps the candidate with the highest proxy accuracy, which uses exactly the per-round budget of FedReason's programs, and runs in every setting.

FedReason uses Claude Opus~5.5~\cite{anthropic2026opus55} (\code{claude-opus-5-5}) with adaptive thinking and medium reasoning effort, the provider's default; the model accepts no temperature or other sampling parameter~\cite{anthropic2026opus55docs}. With $T=10$ the LLM is called at rounds 1, 11, 21, 31, 41 and 51. In the width-heterogeneous settings FedReason uses HeteroFL's static extraction, so that its comparison with HeteroFL isolates aggregation. Four ablations run on CIFAR-10 with $\alpha=0.1$ and seed~0. The metrics-only variant removes the weights and all weight-level helpers from the sandbox and from the programs; it keeps the metadata, the history, the proxy evaluations and weighted averages of client models. The no-proxy variant sets both proxy budgets to zero and removes the proxy metrics from the history. The other two variants call the LLM every five rounds or use high reasoning effort.

We report the test accuracy of the global model after the last round as mean $\pm$ standard deviation over seeds 0, 1 and 2 in the label-skew settings and seeds 0 and 1 elsewhere, rounded half up from the exact means. Under the same seed, all methods see the same partition, initial model and malicious clients. Comparisons are paired by setting and seed. For each baseline we report the mean difference $\Delta$ to FedReason in percentage points (pp) with a 95\% $t$ interval, the wins, ties ($|\Delta|<0.1$) and losses, and an exact two-sided Wilcoxon signed-rank test with Holm's correction over the baselines of a group. With 12 pairs these tests are only indicative, and with four pairs no two-sided test can fall below $p=0.125$.

Under attack we also measure the share $\omega_t$ of the aggregate that goes to the malicious clients in round $t$ and report its mean $\bar\omega$ over the 60 rounds. For the baselines it comes from the coefficients their rule applies. FedReason's programs may clip, rescale or treat layers differently, so for FedReason we use effective weights, the non-negative coefficients
\begin{equation}
  c^t=\arg\min_{c\ge0}\Big\lVert\sum_ic_iU_i^t-\big(G^{t+1}-G^t\big)\Big\rVert_2
  \label{eq:effective}
\end{equation}
that best explain the applied update (they equal $p_i$ for FedAvg). A client counts as excluded in a round if its share is below 0.025, a quarter of the uniform weight. We report the F1 score of the excluded set against the malicious set, averaged over rounds, and the detection round (Det.), from which all malicious clients stay excluded.

\section{Results}
\label{sec:results}

The 168 runs cover the four groups of Table~\ref{tab:settings}. Table~\ref{tab:main} and Fig.~\ref{fig:curves} report label skew, Table~\ref{tab:hetero} width heterogeneity, Table~\ref{tab:robust} and Fig.~\ref{fig:robust} the Byzantine settings, and Table~\ref{tab:ablation} the ablations. The two figures plot the mean over seeds for a subset of the methods, with a band of $\pm1$ standard deviation for FedReason only. In all 24 FedReason runs, every one of the 150 invocations ended with a program that passed its dry run at the first submission, and no round fell back to FedAvg on the server.

\subsection{Statistical and model heterogeneity}
\label{sec:res-het}

\begin{table}[t]
\centering
\caption{Final test accuracy (\%) under label skew. Bold: best; underlined: best server-only rule; $\dagger$: changes clients.}
\label{tab:main}
\small
\begin{tabular}{@{}lccccc@{}}
\toprule
 & \multicolumn{2}{c}{CIFAR-10} & \multicolumn{2}{c}{CIFAR-100} & \\
\cmidrule(lr){2-3}\cmidrule(lr){4-5}
Method & $\alpha=0.1$ & $\alpha=0.5$ & $\alpha=0.1$ & $\alpha=0.5$ & Mean \\
\midrule
FedAvg & 65.72\pmstd{4.80} & 87.34\pmstd{0.34} & 59.95\pmstd{1.07} & 66.24\pmstd{0.17} & 69.81 \\
FedProx$^{\dagger}$ & 67.65\pmstd{4.18} & 87.14\pmstd{0.57} & 59.03\pmstd{0.91} & 65.90\pmstd{0.75} & 69.93 \\
FedAvgM & \underline{67.18}\pmstd{2.43} & \underline{87.87}\pmstd{0.69} & 59.68\pmstd{1.33} & 65.00\pmstd{0.14} & 69.93 \\
FedAdam & 66.59\pmstd{3.49} & 87.13\pmstd{0.43} & 59.23\pmstd{0.76} & 65.41\pmstd{0.17} & 69.59 \\
FedNova & 46.59\pmstd{15.51} & 87.23\pmstd{1.46} & 59.88\pmstd{1.15} & 66.16\pmstd{0.75} & 64.97 \\
SCAFFOLD$^{\dagger}$ & \textbf{70.52}\pmstd{2.72} & \textbf{88.22}\pmstd{0.37} & \textbf{62.39}\pmstd{0.88} & \textbf{68.09}\pmstd{0.52} & \textbf{72.31} \\
FedAdp & 58.61\pmstd{7.07} & 85.38\pmstd{1.86} & 59.64\pmstd{1.37} & 66.20\pmstd{0.23} & 67.46 \\
ProxySel & 65.69\pmstd{5.60} & 87.31\pmstd{0.01} & 60.26\pmstd{0.88} & 66.70\pmstd{0.08} & 69.99 \\
FedReason & 66.53\pmstd{6.34} & 87.77\pmstd{0.95} & \underline{62.26}\pmstd{1.35} & \underline{67.72}\pmstd{0.88} & \underline{71.07} \\
\bottomrule
\end{tabular}
\end{table}

\begin{figure}[!t]
  \centering
  \includegraphics[width=\textwidth]{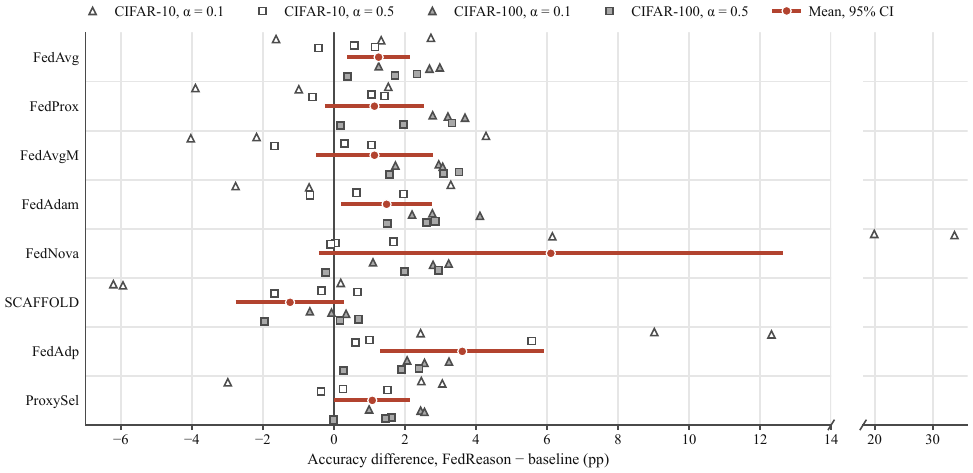}
  \caption{Paired accuracy difference to each baseline under label skew; bar: mean and 95\% CI.}
  \label{fig:paired}
\end{figure}

Under label skew FedReason was more accurate than FedAvg in all four settings (Table~\ref{tab:main}), by 0.81~pp on CIFAR-10 with $\alpha=0.1$, 0.43~pp with $\alpha=0.5$, 2.31~pp on CIFAR-100 with $\alpha=0.1$ and 1.48~pp with $\alpha=0.5$. Over the 12 setting and seed pairs the mean gain was 1.26~pp (95\% CI 0.38 to 2.14), with 10 wins and 2 losses (Wilcoxon $p=0.016$, 0.097 after Holm's correction). Against ProxySel, which spends the same per-round proxy budget on choosing among three standard rules, the mean gain was 1.08~pp (95\% CI 0.01 to 2.15) with 9 wins, 1 tie and 2 losses ($p=0.064$). FedReason also beat FedAdp in all 12 pairs (+3.61~pp, Holm-adjusted $p=0.004$). Its advantage was largest on CIFAR-100, where it was the most accurate rule that changes only the server. In the six CIFAR-100 pairs it beat FedAvg, FedAvgM, FedAdam and FedAdp every time, FedNova five times, and ProxySel five times with one tie. On CIFAR-10 it was not the best server-side rule: FedAvgM had the higher mean with both concentrations (67.18\% against 66.53\% and 87.87\% against 87.77\%), and the pooled difference to FedAvgM (+1.14~pp, 9 wins in 12) comes from CIFAR-100. Fig.~\ref{fig:paired} shows the 12 pairs behind each of these comparisons. On CIFAR-100 the only losses were two to SCAFFOLD and one of 0.23~pp to FedNova. Every loss larger than 2~pp came from CIFAR-10 with $\alpha=0.1$, and so did FedNova's two largest differences (19.92 and 33.76~pp), which make its interval wide.

SCAFFOLD was the most accurate method in all four settings. FedReason trailed it by 1.24~pp on average (95\% CI $-2.77$ to $+0.29$; 5 wins, 1 tie, 6 losses), by 3.99~pp on CIFAR-10 with $\alpha=0.1$ and by less than 0.5~pp in the other three settings. SCAFFOLD changes the clients: each keeps a control variate, corrects every local step with it and sends a second vector to the server. FedReason changes only the server, so the two act at different points and could be combined, which we have not tried.

\begin{figure}[t]
  \centering
  \includegraphics[width=\textwidth]{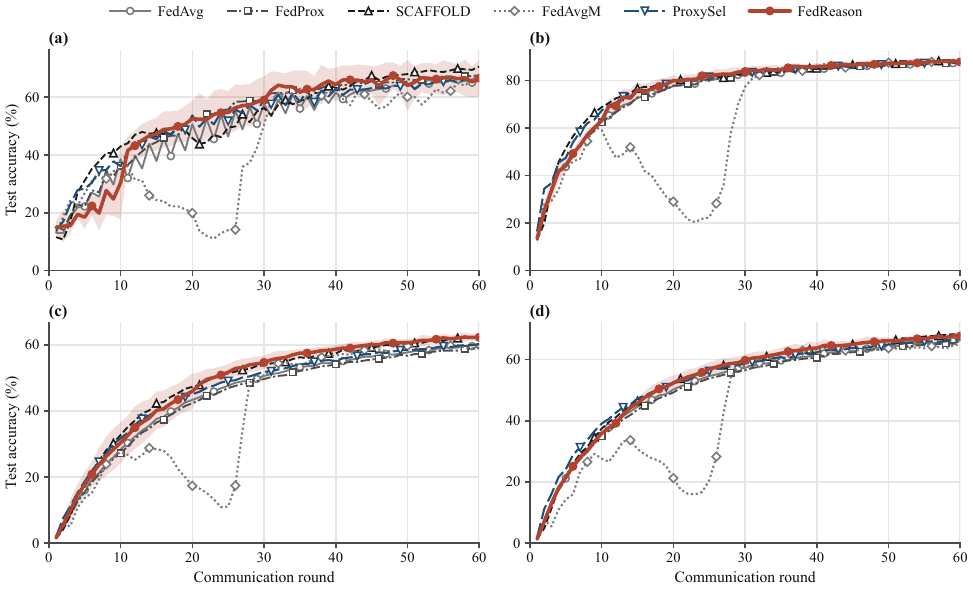}
  \caption{Test accuracy under label skew. Rows: CIFAR-10, CIFAR-100; columns: $\alpha=0.1$, $0.5$.}
  \label{fig:curves}
\end{figure}

The spread across seeds was large for every method on CIFAR-10 with $\alpha=0.1$. FedReason reached 73.84\%, 62.53\% and 63.23\% on the three seeds, and FedAvg 71.11\%, 64.16\% and 61.90\%. A seed changes the partition, the initial model and the LLM's decisions together, so three seeds reveal only large effects. These partitions are also strongly unbalanced in size, with 12 to 264 local steps per client and round across seeds. FedNova normalises each update by its number of local steps, and its low mean in this setting comes mainly from one seed (28.77\%). The curves in Fig.~\ref{fig:curves} show a temporary collapse of FedAvgM in all four settings: its mean accuracy falls to between 11\% and 20\% around rounds 20 to 25 and recovers by round 30. FedReason's mean curve on CIFAR-10 with $\alpha=0.1$ lags in the first ten rounds because of one seed, which Section~\ref{sec:res-llm} explains.

\begin{figure}[!t]
  \centering
  \includegraphics[width=\textwidth]{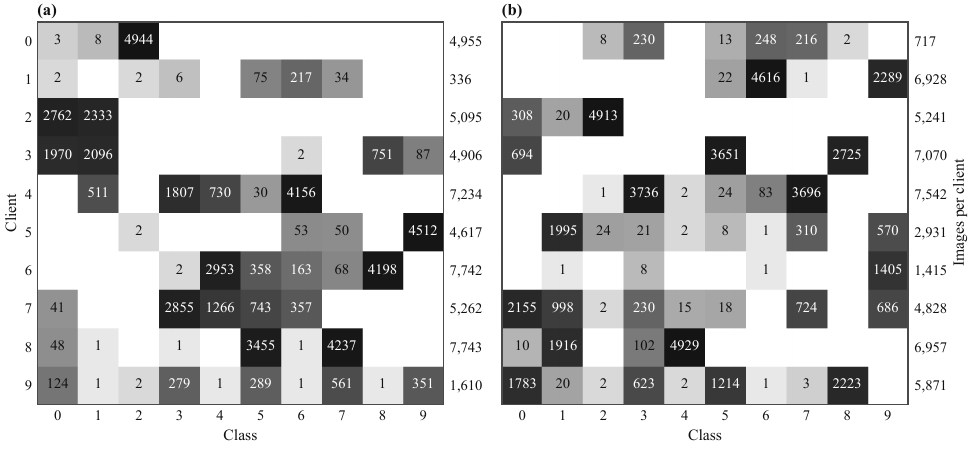}
  \caption{Client label counts on CIFAR-10 with $\alpha=0.1$: (a) seed~0, (b) seed~1.}
  \label{fig:partitions}
\end{figure}

Fig.~\ref{fig:partitions} shows the partitions of seeds 0 and 1 in this setting. In seed~0 one client holds almost all images of class~2 and little else. Seed~1 has two such classes, class~2 at client~2 and class~4 at client~8, but both clients hold other classes as well, and client~6 holds almost only class~9, a class that three other clients share. This is also the seed in which FedReason lost to every baseline except FedNova and FedAdp (Fig.~\ref{fig:paired}). Both runs started with a program that cut a client's weight to a fifth when its leave-one-out cosine was repeatedly negative. The rule never fired in seed~0. In seed~1 every later program kept it, and it fired on two of these honest clients, on client~6 in 11 rounds and on client~2 in four. At round~21 the LLM called client~6 ``the honest single-class-9 holder'' but did not change the rule. How much accuracy this cost is unclear. Over rounds 51 to 60 FedReason averaged 63.25\% in seed~1 against 62.86\% for FedAvg, so its final loss of 1.63~pp to FedAvg lies within the swing from round to round. SCAFFOLD's lead is larger: it averaged 67.96\% over the same rounds.

\begin{table}[t]
\centering
\caption{Final test accuracy (\%) of each sub-model width (two seeds). Bold: best.}
\label{tab:hetero}
\small
\setlength{\tabcolsep}{4pt}
\begin{tabular}{@{}lcccccccc@{}}
\toprule
 & \multicolumn{4}{c}{CIFAR-10} & \multicolumn{4}{c}{CIFAR-100} \\
\cmidrule(lr){2-5}\cmidrule(lr){6-9}
Method & $\times1$ & $\times\frac12$ & $\times\frac14$ & Mean & $\times1$ & $\times\frac12$ & $\times\frac14$ & Mean \\
\midrule
FedAvg & 74.94\pmstd{0.40} & 74.91\pmstd{5.93} & 71.21\pmstd{6.05} & 73.69 & 46.07\pmstd{0.62} & 47.34\pmstd{0.19} & 43.82\pmstd{0.55} & 45.74 \\
HeteroFL & 75.36\pmstd{0.44} & \textbf{75.62}\pmstd{6.12} & \textbf{73.24}\pmstd{2.43} & \textbf{74.74} & 46.33\pmstd{0.95} & 47.82\pmstd{0.48} & 44.49\pmstd{1.35} & 46.21 \\
FedRolex & \textbf{77.29}\pmstd{1.21} & 59.91\pmstd{4.81} & 26.08\pmstd{0.62} & 54.42 & 47.17\pmstd{0.54} & 29.18\pmstd{0.83} & 6.24\pmstd{1.68} & 27.53 \\
ProxySel & 75.19\pmstd{0.11} & 74.82\pmstd{7.18} & 72.83\pmstd{5.32} & 74.28 & 44.03\pmstd{0.21} & 46.23\pmstd{0.04} & 44.06\pmstd{1.19} & 44.77 \\
FedReason & 76.39\pmstd{0.07} & 66.46\pmstd{17.92} & 70.76\pmstd{4.60} & 71.20 & \textbf{47.73}\pmstd{0.59} & \textbf{48.63}\pmstd{0.45} & \textbf{46.14}\pmstd{0.62} & \textbf{47.50} \\
\bottomrule
\end{tabular}
\end{table}

\begin{figure}[!t]
  \centering
  \includegraphics[width=\textwidth]{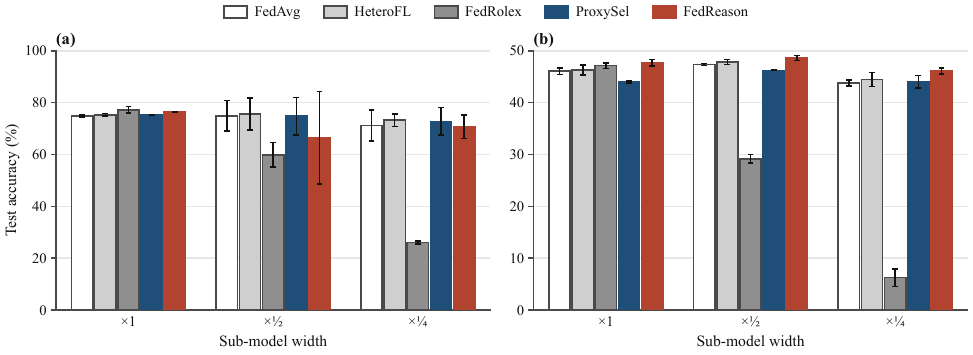}
  \caption{Test accuracy per sub-model width (mean and std, two seeds): (a) CIFAR-10, (b) CIFAR-100.}
  \label{fig:hetero}
\end{figure}

With sub-models of different widths (Table~\ref{tab:hetero} and Fig.~\ref{fig:hetero}), FedReason was the most accurate method at every width on CIFAR-100 (47.73\%, 48.63\% and 46.14\% for $\times1$, $\times\frac12$ and $\times\frac14$). Its mean over widths was 47.50\% against 46.21\% for HeteroFL, which uses the same extraction. On CIFAR-10 the result is mixed. For the global model FedRolex was best (77.29\%), followed by FedReason (76.39\%) and HeteroFL (75.36\%). FedRolex's narrow sub-models, however, collapsed (26.08\% at $\times\frac14$ on CIFAR-10 and 6.24\% on CIFAR-100), as one may expect from an extraction that trains a moving window rather than the leading channels we evaluate. FedReason's $\times\frac12$ sub-model was unstable on CIFAR-10, with 53.79\% and 79.13\% in the two seeds (the long error bar in Fig.~\ref{fig:hetero}a), so its mean over widths (71.20\%) stayed below HeteroFL's (74.74\%). Section~\ref{sec:res-llm} traces the bad seed to the exclusion of two honest $\times\frac12$ clients. Pooled over both datasets, FedReason's global model was 1.22~pp more accurate than HeteroFL's and won all four pairs ($p=0.125$, the smallest value that four pairs allow).

\subsection{Byzantine clients}
\label{sec:res-byz}

\begin{table}[t]
\centering
\caption{Accuracy (\%) and attacker influence with 3 of 10 clients malicious (CIFAR-10, $\alpha=0.5$). Bold: best.}
\label{tab:robust}
\small
\setlength{\tabcolsep}{4pt}
\begin{tabular}{@{}lcccccccc@{}}
\toprule
 & Clean & \multicolumn{3}{c}{Sign flip} & \multicolumn{4}{c}{Label flip} \\
\cmidrule(lr){2-2}\cmidrule(lr){3-5}\cmidrule(lr){6-9}
Method & Acc. & Acc. & $\bar\omega$ (\%) & F1 & Acc. & $\bar\omega$ (\%) & F1 & Det. \\
\midrule
FedAvg & 87.34\pmstd{0.34} & 10.00\pmstd{0.00} & 29.11 & 0.00 & 78.93\pmstd{5.47} & 29.11 & 0.00 & never \\
Median & 84.94\pmstd{4.89} & 57.17\pmstd{2.11} & 9.65 & 0.00 & 71.19\pmstd{0.49} & 26.68 & 0.00 & never \\
Trimmed mean & 86.07\pmstd{2.61} & 51.44\pmstd{5.88} & 10.49 & 0.00 & 71.94\pmstd{0.08} & 26.98 & 0.00 & never \\
Krum & 52.95\pmstd{11.91} & 51.82\pmstd{10.30} & \textbf{0.00} & 0.50 & 52.70\pmstd{11.55} & \textbf{0.83} & 0.50 & \textbf{1.5} \\
FLTrust & \textbf{87.81}\pmstd{0.06} & \textbf{85.95}\pmstd{0.89} & \textbf{0.00} & \textbf{1.00} & \textbf{86.20}\pmstd{0.71} & 2.44 & \textbf{0.89} & 32.5 \\
ProxySel & 87.31\pmstd{0.01} & 10.00\pmstd{0.00} & 29.11 & 0.00 & 80.47\pmstd{4.89} & 29.11 & 0.00 & never \\
FedReason & 87.77\pmstd{0.95} & 85.24\pmstd{0.97} & 0.43 & \textbf{1.00} & 84.94\pmstd{0.14} & 4.31 & 0.85 & 16 \\
\bottomrule
\end{tabular}
\end{table}

Under sign flipping the three malicious updates dominate the average, and FedAvg and ProxySel stayed at chance level (Table~\ref{tab:robust}). The median, the trimmed mean and Krum reached 51\% to 57\%. FLTrust (85.95\%) and FedReason (85.24\%) were the only methods that came close to the attack-free accuracy of about 87.8\%. Under label flipping FedAvg kept 78.93\%, the median and the trimmed mean fell to 71\% to 72\% and Krum to 52.70\%, while FLTrust reached 86.20\% and FedReason 84.94\%. Over the four attack pairs FedReason was 0.99~pp below FLTrust (95\% CI 0.05 to 1.92~pp) and lost all four. The two methods use the proxy set differently. FLTrust trains a root model on it in every round and scores the clients against that update, whereas FedReason never trains on the server and uses the proxy set only for forward evaluations. Without attack the two were level (87.81\% and 87.77\%; FedReason has three seeds there and the robust rules two). Over the same two seeds, the median lost 2.38~pp and Krum 34.37~pp relative to FedAvg.

\begin{figure}[t]
  \centering
  \includegraphics[width=\textwidth]{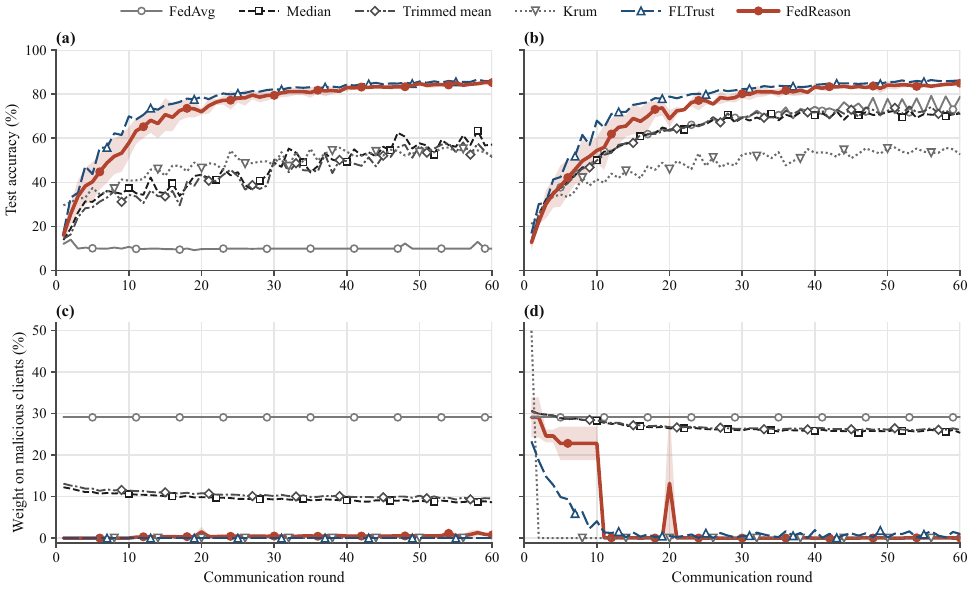}
  \caption{Test accuracy (top) and weight on malicious clients (bottom); left: sign flipping, right: label flipping.}
  \label{fig:robust}
\end{figure}

FedReason gave the sign-flipping clients 0.43\% of the aggregate on average and excluded all three from round~1 in both seeds, without excluding any honest client (F1 of 1.00; in Fig.~\ref{fig:robust}c the zero lines of Krum and FLTrust lie over the FedReason line). The residual share comes from the fit of Eq.~\eqref{eq:effective}, not from the programs. From round~11 the programs aggregated the classifier rows class by class, and no mixture of client updates reproduces such an update exactly (relative residual about 0.2 to 0.3). The fit then spread a little weight onto the attackers, whereas the coefficients the programs reported gave them zero weight in all rounds but one. The round-1 transcripts show how FedReason decided. In seed~0 the LLM found three updates with about three times the median norm, a positive cosine among themselves and a cosine of about $-0.2$ with the mean of the other seven, and called this a coordinated attack. On the proxy set, excluding the three scored 12.6\% against 11.4\% for FedAvg, a difference that the LLM itself described as round-1 noise, so the decision rested on the weights. The three large updates dominated every leave-one-out mean, and \code{summary()} therefore flagged all seven honest clients for a negative leave-one-out cosine. The LLM used the median pairwise cosine as reference instead; \ref{app:program} lists the resulting program.

Label flipping was harder to see. In seed~0 the round-1 program excluded one attacker as a norm outlier from round~5, and at round~11 the LLM excluded the other two. It cited a cluster of the three in the pairwise cosines and classifier updates of 34\% to 46\% of the classifier norm, against 18\% to 27\% for the other clients. On the proxy set the three client models scored 5.6\%, 7.6\% and 13.4\% on their own, against 37.6\% for an honest but skewed client. In seed~1 the round-11 invocation excluded all three attackers on similar evidence, but the program readmitted them in round~20 (the spike in Fig.~\ref{fig:robust}d) before the next invocation excluded them for good. All three attackers thus stayed excluded from round~11 in seed~0 and from round~21 in seed~1 (Det.\ 16 in Table~\ref{tab:robust}), and their share of the aggregate over the last 30 rounds was 0.02\%. The F1 score over all rounds was 0.85, against 0.89 for FLTrust, whose detection round was 32.5. Krum detects the attackers early only because it keeps a single client and so excludes six honest clients per round. FedReason excluded an honest client in two rounds of seed~1, which the LLM noticed and corrected at round~31. Because the label-flip identification used proxy evaluations of single client models, it does not rest on the weights alone.

\subsection{Ablations and LLM behaviour}
\label{sec:res-llm}

\begin{table}[t]
\centering
\caption{Ablations on CIFAR-10, $\alpha=0.1$, seed~0.}
\label{tab:ablation}
\small
\begin{tabular}{@{}lccccc@{}}
\toprule
Variant & Acc. (\%) & $\Delta$ (pp) & Proxy evaluations & Requests per invocation & Cost per run (USD) \\
\midrule
FedAvg & 71.11 & $-2.73$ & 0 & & \\
ProxySel & 71.38 & $-2.46$ & 180 & & \\
FedReason & 73.84 & ref. & 169 & 6.3 & 1.22 \\
\quad high effort & 73.87 & $+0.03$ & 114 & 7.3 & 1.80 \\
\quad every 5 rounds & 70.16 & $-3.68$ & 138 & 4.3 & 1.78 \\
\quad metrics only & 67.86 & $-5.98$ & 163 & 2.5 & 0.48 \\
\quad no proxy feedback & 38.84 & $-35.00$ & 0 & 2.3 & 0.55 \\
\bottomrule
\end{tabular}
\end{table}

\begin{figure}[!t]
  \centering
  \includegraphics[width=\textwidth]{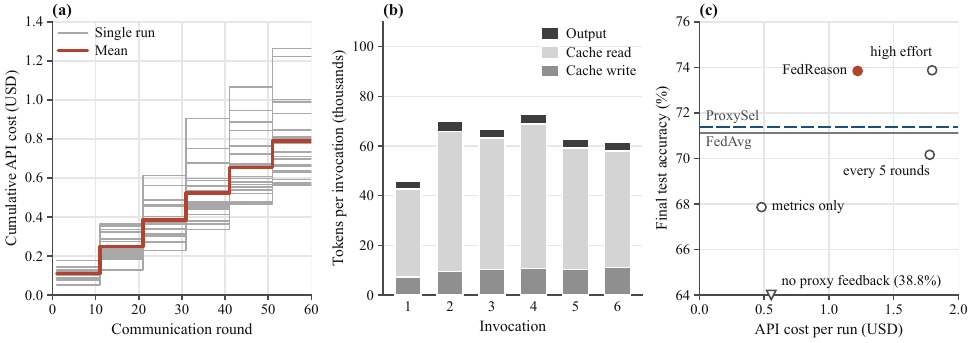}
  \caption{API cost: (a) cumulative per run; (b) tokens per invocation; (c) ablation accuracy against cost.}
  \label{fig:cost}
\end{figure}

Table~\ref{tab:ablation} lists the ablations, all on CIFAR-10 with $\alpha=0.1$ and seed~0, where full FedReason reached 73.84\% against 71.11\% for FedAvg and 71.38\% for ProxySel. With one seed per variant and an 11-point spread between the seeds of the full method in this setting, only large differences carry information. High reasoning effort changed accuracy by 0.03~pp and raised the cost per run from 1.22 to 1.80~USD. Calling the LLM every five rounds gave 70.16\% at 1.78~USD, so more frequent invocations did not help in this seed. In Fig.~\ref{fig:cost}c only the full method and the high-effort variant lie above both FedAvg and ProxySel.

Without the weights, accuracy fell to 67.86\%. The metrics-only LLM still found a way to probe the clients: at rounds 31 and 41 it passed one-hot coefficients to the weighted-average helper, evaluated single client models on the proxy set and inferred each client's classes from the per-class accuracies. This variant therefore removes weight-level analysis rather than every signal derived from the weights, and the gap of about 6~pp, in this single seed, is our only estimate of what that analysis adds over black-box probing.

Removing proxy feedback did far more damage, and accuracy dropped to 38.84\%. The LLM again recognised the label skew and identified the clients that alone held a class. At round~1, however, it adopted class-evidence weighting of the classifier rows, the heuristic of the example program in the system prompt, and kept it at every later invocation. Its rationales give the reason: ``Without proxy feedback I avoid untested changes such as server momentum.'' Test accuracy stayed at 10\% to 13\% in odd rounds and between 10\% and 32\% in even rounds up to round~40, and at round~51 the LLM resubmitted what it called ``the proven policy'' for the last ten rounds. The same heuristic stalled seed~2 of the full method. Its round-1 program held proxy accuracy between 13\% and 23\% for ten rounds while the global model predicted almost only one class. At round~11 the LLM saw the stagnation in the proxy history, traced it to the classifier weighting and removed it; this is the lag in Fig.~\ref{fig:curves}a. The seed-0 run tested the same weighting at four invocations and never adopted it. The weights alone did not reveal the harm, and a few proxy evaluations did. Part of this effect is due to the prompt, since the harmful heuristic was the one its example program suggested.

\begin{figure}[t]
  \centering
  \includegraphics[width=\textwidth]{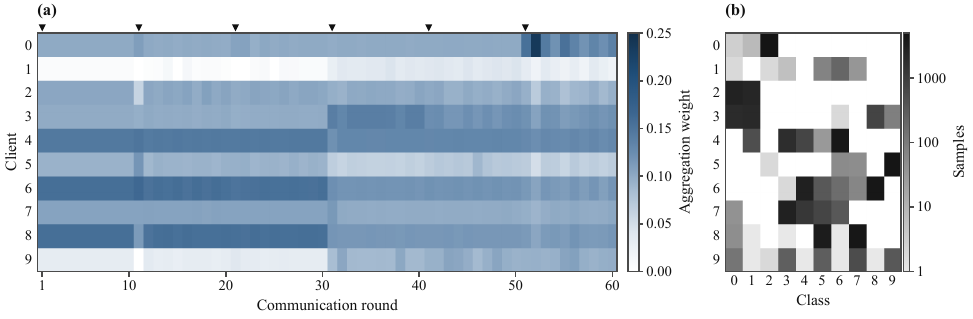}
  \caption{Seed-0 run on CIFAR-10, $\alpha=0.1$: (a) effective weights (triangles: LLM calls); (b) client label counts.}
  \label{fig:weights}
\end{figure}

Fig.~\ref{fig:weights} follows the seed-0 run; panel~(b) shows the client label counts, which the LLM never saw. The first program was a conservative variant of FedAvg, with sample-count weights, clipping at twice the median update norm and a slow suspicion counter, so the two smallest clients (336 and 1{,}610 images) carried as little weight as under FedAvg. In this seed FedAvg alternated between rounds, with 35.5\%, 41.8\%, 36.7\% and 47.8\% in rounds 11 to 14. At round~11 the LLM found classifier updates of 44\% to 118\% of the classifier's norm and per-class proxy accuracies that swung from round to round. It measured a proxy accuracy of 50\% for a halved classifier step against 36\% for FedAvg and submitted a program that damps the classifier update, after which accuracy in rounds 11 to 14 was 47.1\%, 49.2\%, 51.6\% and 52.1\%. At round~31 it mixed the sample weights with class-balanced weights estimated from the classifier-bias changes, which moved weight towards clients with rare classes (the change at round~31 in Fig.~\ref{fig:weights}a). The proxy gain was small (68.4\% against 67.6\%), in its own words ``within noise but principled''. Its statements about single clients agree with the hidden partition. It described client~0 as the only holder of class~2 and client~5 as a single-class client holding class~9. The partition confirms both: client~0 holds 4{,}944 of its 4{,}955 images in class~2, no other client holds more than two images of that class, and client~5 holds 4{,}512 of its 4{,}617 images in class~9. Its estimates for clients with several classes changed between invocations, which the LLM itself put down to noise. From round~51 the program also tried bias offsets of 0, 0.8 and 1.6 for the weakest class in every round and kept the one that scored best on the proxy set, within its three evaluations; nothing was fitted by gradient descent. No mixture of client models reproduces such a shift, so in rounds 51 to 60 the effective weights explain the applied update only roughly (relative residual up to 0.88, against about 0.05 before).

Across the 120 programs accepted in the 20 main runs, the LLM settled on a consistent set of tools. Almost all programs clipped update norms (98\%) and down-weighted clients whose update direction disagreed with the others (91\%), and every program kept per-client suspicion counters in \code{ctx.memory} and returned FedAvg from an exception handler. Class-aware aggregation of the classifier rows appeared in 38\% of the programs, a server step other than one in 22\% and server momentum in 35\%, never in the Byzantine runs. Width-aware weighting appeared in 92\% of the programs for the width-heterogeneous settings, and selection among candidates on the proxy set in 45\%. The programs used 0.95 of their three evaluations per round on average; including the analysis evaluations, a main run requested 74 candidate evaluations, against 180 for ProxySel. No program aggregated with a coordinate-wise median or trimmed mean, and one run used the median only as a reference direction for cosine scores. In the 12 label-skew runs, no program gave any client zero weight, whereas the Byzantine programs excluded clients in 225 of 240 rounds. In 28\% of the invocations after the first, the LLM resubmitted the previous program unchanged. On average a program's output differed from FedAvg's by 0.40 times the norm of the FedAvg update under label skew and 0.82 times under attack.

The audit turned up one programming error, which the LLM caught and fixed on its own. On CIFAR-100 with $\alpha=0.1$ (seed~0), the program submitted at round~11 multiplied tensors of incompatible shapes in its classifier weighting. Its own exception handler caught the error and fell back to a sample-weighted classifier update, so the server recorded no failure. The error message appeared in the program's printed diagnostics, which the next brief contains, and at round~21 the LLM wrote that ``policy\_011 had a broadcasting bug, so its head always fell back to plain FedAvg'' and submitted a corrected program. Three further revisions concerned heuristics that had misfired in width-heterogeneous runs, where leave-one-out cosines penalise clients for their width rather than for their data. On CIFAR-100 the LLM noticed at round~11 that its rule had pushed the weights of the three full-width clients down to about 0.02, called it ``an artifact of heterogeneous widths'' and replaced the rule.

Not every misfire was caught. In the CIFAR-10 width-heterogeneous run with seed~0, the LLM read strongly negative class-9 bias changes of clients 3 and 5 (both $\times\frac12$) as label-flip poisoning (``That fits a 9-flip poisoning'', round~51). It excluded them in 36 of the 40 rounds from round~21 on. This setting has no attackers, and neither client holds a single image of class~9, which is exactly when honest training pushes a class bias down. The exclusion raised the proxy accuracy of class~9, but the proxy set evaluates only the full model and cannot see the $\times\frac12$ sub-model. That sub-model reached 53.7\%, 52.5\% and 53.8\% at rounds 30, 50 and 60, when the two clients were excluded, against 58.5\% and 63.4\% at rounds 20 and 40, when they were not. This explains the large spread of the $\times\frac12$ result in Table~\ref{tab:hetero}.

An invocation took $3.9\pm1.5$ model requests (2 to 9), 2.9 sandbox calls and 2.9 analysis evaluations. It read about 60{,}000 input tokens, 83\% of them from the prompt cache, and wrote about 3{,}600 output tokens including thinking. It cost $0.13\pm0.05$~USD and took $41\pm18$~s of wall-clock time (median 36~s). A 60-round run with six invocations cost 0.79~USD on average (0.56 to 1.26~USD). Panels (a) and (b) of Fig.~\ref{fig:cost} show how this cost accrues over a run and how the tokens of an invocation divide into cache reads, cache writes and output. The first invocation was the smallest, with about 46{,}000 tokens against 62{,}000 to 73{,}000 for the later ones. The 24 FedReason runs, ablations included, cost 20.38~USD; the API bill of the study was 20.51~USD, because it also contains one aborted first start of a run. In our simulation a FedAvg round took about 16~s, so 60 rounds took 16~min. The six invocations of a FedReason run took about 4~min in total, roughly a quarter of that, and executing the programs added about 1.6~s per round, or 1.5~min per run. The measured wall time of a FedReason run was longer (37~min on average), but most of the difference lies in client training, which is identical for both methods and was slowed by other runs that shared the GPU. In cross-silo deployments, where rounds take minutes or longer, the relative delay of the invocations is smaller.

\section{Discussion and limitations}
\label{sec:discussion}

The results support a modest claim. With changes only on the server and a small budget of forward evaluations, an LLM that reads the client models wrote programs that beat FedAvg in most paired label-skew comparisons and beat a control with the same per-round proxy budget in most of them. It also identified sign-flipping clients from their weights in the first round. The gains under label skew were about one point and largest on CIFAR-100. FedReason did not beat SCAFFOLD, did not beat FLTrust under attack, and was worse than HeteroFL on CIFAR-10 sub-models because of one wrong inference. Our experiments contain no change of regime during training, since the partitions are fixed and the attackers act from the first round, so the motivation that a program can follow a changing situation remains untested.

The proxy set matters more than the weights alone. Without it the LLM held on to a harmful heuristic for a whole run, while the full method dropped the same heuristic after ten rounds. Our proxy set is the most favourable kind: a class-balanced sample of the training split and therefore of the test distribution, although small (5 images per class on CIFAR-100). FLTrust and ProxySel make the same assumption. A biased, mislabelled or class-imbalanced proxy set could mislead all three, and we did not study this. ProxySel shows that proxy access used as selection among three fixed rules does not explain FedReason's gains, but FedReason used its budget more freely, for example to evaluate single client models, which a fixed control does not do.

The rationales and notes are logs written by the model, not verified explanations. In the cases we checked against information the LLM never saw, most claims were correct, for example the single-class clients and the identities of the sign-flipping clients. Others were not: the label-flip diagnosis of two honest clients in a width-heterogeneous run, and estimates of label composition that changed between invocations. Exclusions should therefore be treated as hypotheses that need corroboration, every served sub-model width should be evaluated, and audits should read the executed code rather than only the rationales. The authors wrote the system prompt knowing the three kinds of heterogeneity we evaluate. Its heuristics and example program shape what the LLM tries first, as the class-evidence weighting showed, whereas the baselines use the hyper-parameters of their original papers without tuning. Part of FedReason's gains may come from domain knowledge in the prompt rather than from the analysis of the weights, and the metrics-only ablation, which keeps the prompt but withholds the weights, separates the two only in part.

The model accepts no sampling parameters, so its outputs cannot be made deterministic, and a provider may change or retire a model version. Three seeds expose only large effects, and the ablations rest on a single seed each. Every accepted program is logged with the rounds in which it was active, so the decisions could be replayed without the LLM, but we have not tested such a replay, and GPU training with reduced precision is not guaranteed to reproduce the same trajectory.

FedReason shares per-client statistics with the model provider. In our transcripts the LLM routinely stated the label composition of individual clients, and it carried these statements forward in its notes. The output limit bounds the volume of what is disclosed but not its content. The method needs individual updates and so cannot be combined with secure aggregation. We have not evaluated two obvious mitigations: filtering or adding noise to the sandbox outputs, and running an open-weight model on premises so that nothing leaves the server. Executing LLM-written code on the server that holds every client model also requires isolation beyond a restricted Python namespace. The attacks we evaluated are simple and non-adaptive. Attackers who know the meta-aggregator and shape their updates to pass its norm, cosine and classifier checks, as adaptive attacks do against fixed robust rules~\cite{fang2020local}, were not tested.

The evaluation uses ten clients with full participation, two image datasets, one architecture, 60 rounds and one LLM. Large cross-device populations with partial participation, larger models and other modalities may change both the benefits and the costs. With thousands of clients per round the sandbox could not expose every client model, and the analysis would have to work on sketches or on summaries of client clusters.

\section{Conclusion}
\label{sec:conclusion}

FedReason replaces the fixed aggregation formula of FL with programs that an LLM writes after reading the client models on the server. The LLM never trains on the server, and every aggregate is checked and logged. In 168 runs on CIFAR-10 and CIFAR-100 with label skew, sub-models of different widths and Byzantine clients, FedReason improved on FedAvg by 1.26~pp on average under label skew and was the strongest server-side rule on CIFAR-100, but it did not beat SCAFFOLD. It identified sign-flipping clients from their weights in the first round and stayed within about one point of FLTrust without training on the server, at 0.79~USD per run. In the ablations, removing proxy feedback cost 35 points and withholding the weights about six, each in one seed, while the transcripts show that the LLM's inferences can be wrong. Within these limits, an LLM that reads weights and writes code, as weight-space programming suggests~\cite{takehi2026sparks}, is a workable part of an FL server: its cost is fixed in advance and each of its decisions can be read afterwards. Open questions include on-premises open-weight models, privacy accounting for the statistics the LLM reads, combination with client-side methods such as SCAFFOLD, and larger client populations.

\appendix

\section{An aggregation program written by the LLM}
\label{app:program}

Listing~\ref{lst:policy} reproduces, verbatim, the first program accepted in the sign-flip run with seed~0, submitted at round~1. It builds the Gram matrix of the client updates and flags a client whose update norm exceeds three times the median norm, or whose median cosine with the other clients is negative and at least 0.05 below the median over all clients. It counts flags per client in \code{ctx.memory} and excludes a client that is flagged in the current round, or in at least three rounds and at least half of all rounds. If fewer than three clients would remain, it relaxes the rule. It then clips the accepted updates to their median norm, averages the normalisation statistics over the accepted clients and returns FedAvg if anything raises an exception. The program aggregated rounds 1 to 10. Its successors kept the exclusion rule and added class-evidence weighting of the classifier rows (round~11), a proxy-checked choice between server steps of 1.0 and 1.3 (round~31) and a proxy-checked moving average of global models for the last rounds (round~51).

\begin{lstlisting}[float=!t,caption={Program \texttt{policy\_001} of the sign-flip run (seed 0), as submitted at round 1.},label={lst:policy}]
def aggregate(ctx):
    torch, np = ctx.torch, ctx.np
    K = ctx.num_clients
    n = np.array([m['num_samples'] for m in ctx.meta], dtype=float)
    cids = [m['cid'] for m in ctx.meta]
    w0 = n / n.sum()
    try:
        M = ctx.update_matrix()
        Gm = M @ M.T
        nr = Gm.diag().clamp_min(1e-12).sqrt()
        C = (Gm / (nr[:, None] * nr[None, :])).numpy().astype(float)
        np.fill_diagonal(C, np.nan)
        medcos = np.nanmedian(C, axis=1)
        norms = nr.numpy().astype(float)
        mednorm = float(np.median(norms))
        gmed = float(np.median(medcos))
        flag = (norms > 3.0 * mednorm) | ((medcos < 0.0) & (medcos < gmed - 0.05))
        mem = ctx.memory.setdefault('susp', {})
        for i, c in enumerate(cids):
            h = mem.setdefault(str(c), [0, 0])
            h[0] += int(flag[i]); h[1] += 1
        excl = flag.copy()
        for i, c in enumerate(cids):
            h = mem[str(c)]
            if h[0] >= 3 and h[0] >= 0.5 * h[1]:
                excl[i] = True
        if (~excl).sum() < 3:
            excl = norms > 3.0 * mednorm
            if (~excl).sum() < 3:
                excl = np.zeros(K, dtype=bool)
        w = w0.copy(); w[excl] = 0.0
        w = w / w.sum()
        # clip accepted updates to the median norm of accepted clients
        accn = norms[~excl]
        cap = float(np.median(accn))
        scale = np.minimum(1.0, cap / np.maximum(norms, 1e-12))
        coef = torch.as_tensor(w * scale, dtype=torch.float32)
        upd = (coef[:, None] * M).sum(0)
        d = ctx.unflatten(upd)
        bn_keys = [k for k, _, kind, _ in ctx.layers if kind == 'bn_stat']
        new = ctx.weighted_average(w, keys=bn_keys)
        new = dict(new)
        for k, v in d.items():
            new[k] = ctx.G[k] + v
        print(f'excl={[cids[i] for i in range(K) if excl[i]]} medcos={np.round(medcos,2).tolist()}')
        return new, {'weights': w.tolist()}
    except Exception as exc:
        print(f'fallback FedAvg: {exc}')
        return ctx.fedavg(), {'weights': w0.tolist()}
\end{lstlisting}

\section*{CRediT authorship contribution statement}

\textbf{Ali Mohammadi Ruzbahani:} Conceptualization, Methodology, Software, Investigation, Formal analysis, Visualization, Writing (original draft). \textbf{Abbas Yazdinejad:} Validation, Writing (review and editing). \textbf{Hadis Karimipour:} Supervision, Writing (review and editing).

\section*{Declaration of competing interest}

The authors declare that they have no known competing financial interests or personal relationships that could have appeared to influence the work reported in this paper.

\section*{Data availability}

CIFAR-10 and CIFAR-100 are publicly available~\cite{krizhevsky2009learning}. The code, configuration files, per-run logs, LLM transcripts and all aggregation programs produced in this study will be made publicly available upon publication.

\section*{Declaration of generative AI and AI-assisted technologies in the writing process}

During the preparation of this work the authors used Claude (Anthropic) in order to improve the readability of the code and to debug it, to improve the text, to correct grammatical errors, to trim the text, and to improve the visualisation of the figures. After using this tool, the authors reviewed and edited the content as needed and take full responsibility for the content of the publication. Separately, Claude Opus~5.5 is a component of the method studied in this paper: it acts as the meta-aggregator described in Sections~\ref{sec:method} and~\ref{sec:setup}, and all of its interactions during the experiments were logged.

\end{document}